# Tuning charge density wave of kagome metal ScV$_6$Sn$_6$

Changjiang Yi[1,*], Xiaolong Feng[1], Nitesh Kumar[2], Claudia Felser[1] and Chandra Shekhar[1,*]

[1] Max Planck Institute for Chemical Physics of Solids, 01187 Dresden, Germany
[2] S. N. Bose National Centre for Basic Sciences, Salt Lake City, Kolkata 700 106, India

**Abstract**

Compounds with a kagome lattice exhibit intriguing properties and the charge density wave (CDW) adds an additional layer of interest to research on them. In this study, we investigate the temperature and magnetic field dependent electrical properties under a chemical substitution and hydrostatic pressure of ScV$_6$Sn$_6$, a non-magnetic charge density wave (CDW) compound. Substituting 5 % Cr at the V site or applying 1.5 GPa of pressure shifts the CDW to 50 K from 92 K. This shift is attributed to the movement of the imaginary phonon band, as revealed by the phonon dispersion relation. The longitudinal and Hall resistivities respond differently under these stimuli. The magnetoresistance (MR) maintains its quasilinear behavior under pressure, but it becomes quadratic after Cr substitution. The anomalous Hall-like behavior of the parent compound persists up to the respective CDW transition under pressure, after which it sharply declines. In contrast, the longitudinal and Hall resistivities of Cr substituted compounds follow a two-band model and originates from the multi carrier effect. These results clearly highlight the role of phonon contributions in the CDW transition and call for further investigation into the origin of the anomalous Hall-like behavior in the parent compound.

[*] Corresponding author: yi@cpfs.mpg.de, shekhar@cpfs.mpg.de

**Introduction**

Kagome-lattice compounds are known to exhibit unavoidable exotic topological electronic states [1–7]. The discovery of the charge density wave (CDW) in some of these compounds has introduced a captivating dimension to the research [8–10]. In $A$V$_3$Sb$_5$ ($A$= K, Rb, Cs), the CDW coexists with a superconducting ground state [5,11–17]. In antiferromagnetic hexagonal-FeGe, the CDW amplifies the ordered moment, justifying a deeper correlation between CDW and magnetism [9,18,19]. Recently, the CDW in ScV$_6$Sn$_6$ was discovered at 92 K, making it the only known compound in the vast HfFe$_6$Ge$_6$ kagome family with the CDW state. The out-of-plane lattice dynamics of the CDW suggests an unconventional nature [10]. Additionally, the CDW exhibits various microscopic features, such as the critical role of phonons [20–23], substantial spin Berry curvature [24], partial bandgap opening [25,26], and hidden magnetism [27]. It is worth noting that the CDW phase of ScV$_6$Sn$_6$ is very fragile without any additional features of superconductivity and magnetism. The application of either internal chemical pressure through doping or external physical pressure can completely suppress the CDW due to minor modifications in the Sc–Sn and Sn–Sn bonds [26,28,29]. The phase remains robust at lower doping levels [26]. The Fermi surface (FS) comprises of electron-type Fermi pockets except for a tiny hole-type pocket, which remain almost unchanged after the CDW transition [30]. Furthermore, ScV$_6$Sn$_6$ exhibits an anomalous Hall-like behavior [30,31], which cannot be explained by multiple band model, but it rather indicates a hidden magnetism origin [27]. Interestingly, this anomalous Hall-like feature coexists with the CDW phase and dies down upon approaching $T_{CDW}$. This prompts a speculation of a correlation between CDW and anomalous Hall-like feature. This speculation can be tested by performing pressure-dependent magneto-transport measurements since CDW phase seems to rely on loose Sc-Sn-Sn chains which are highly impacted by pressure [26,28,29]. Owing to several exotic microscopic features associated with the Sc–Sn and Sn–Sn bonds, as well as hidden magnetism related to anomalous Hall-like behavior, it is compelling to investigate that how the anomalous Hall-like behavior is coupled with the CDW phase. In this study, we have performed the temperature-dependent electrical transport of ScV$_6$Sn$_6$ under pressure and Cr doping compounds at the V site. This comprehensive study highlights how physical and chemical pressure affect the CDW, emphasizing the changes in electronic properties, such as conductivity, band structure and FS.

**Results and discussion**

The parent compound ScV$_6$Sn$_6$ crystallizes in a hexagonal centrosymmetric structure with a *P*6/*mmm* space group at room temperature. The V atoms form a double kagome lattice and are separated by ScSn$_2$ and Sn$_2$ layers along the *c*-axis (figure 1a). High quality hexagon shaped single crystals were grown using the flux method, see supplemental material (SM) [32]. To investigate the effects of internal chemical and external physical pressure on the CDW phase and related transport properties in ScV$_6$Sn$_6$, we performed systematic measurements on pristine single crystal under hydrostatic pressures and Cr substituted single crystals. The electric current and magnetic field were applied in the *ab*-plane and *c*-axis, respectively, as listed in table S1 in SM [32]. For this study, we have selected two concentrations Sc(V$_{1-x}$Cr$_x$)$_6$Sn$_6$ (nominal $x$ = 0.05, 0.1) among the various levels of Cr substitution, along with the pristine compound. The radius of Cr atoms is similar to that of V atoms, which means that the lattice parameters would not change significantly. Table S2 in SM [32] summarizes the estimated lattice parameters of Cr substituted samples and the pristine crystal, obtained from the refinement of powder x-ray diffraction (PXRD) pattern (figure S1 in SM) [32]. In figure 1a, the Cr occupies the V atomic position. The real substituting components $x$ is approximately 0.086 for Sc(V$_{0.95}$Cr$_{0.05}$)$_6$Sn$_6$ and ~0.157 for Sc(V$_{0.9}$Cr$_{0.1}$)$_6$Sn$_6$, which were averaged from several points measured by Energy-dispersive x-ray spectroscopy (EDXS) (see table S3 in SM) [32]. The observed concentrations dependent $T_{CDW}$ of our samples are consistent with the previous report, where the CDW $T_{CDW}$ is approximately 60 K for $x = 0.06$ [26], but approximately 48 K for $x = 0.086$ in our work. Figures 1b and 1c represent the clear Laue diffraction patterns with six-fold symmetry along the $c$-axis for both crystals, indicating their high quality. Table S1 in SM [32] lists the directions of external magnetic field and applied current used during the electronic transport measurements.

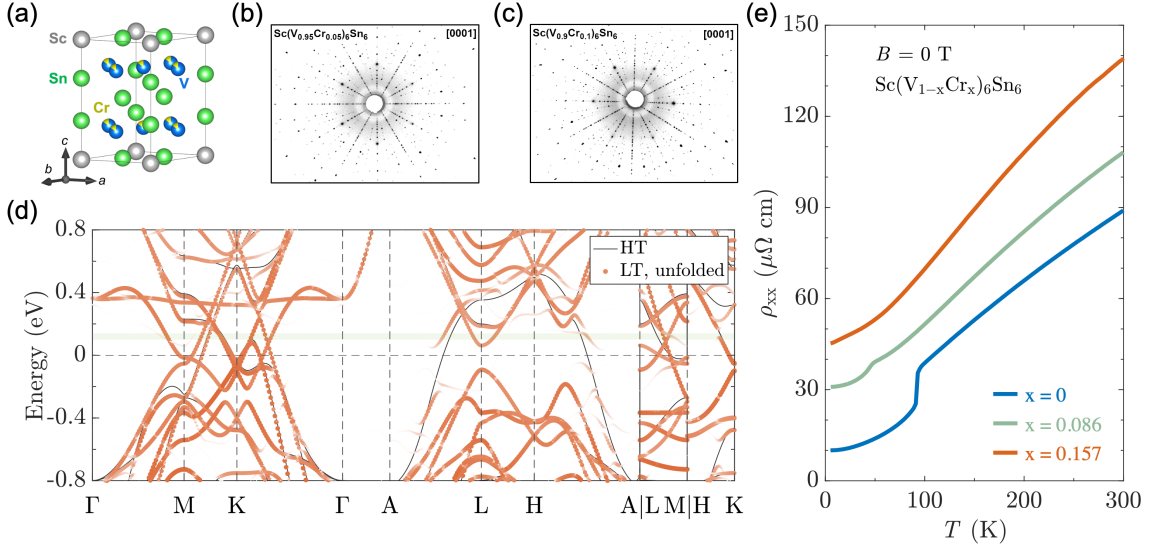

**Figure 1.** Crystal structure, Laue pattern, band structure and resistivity. (a) A representative crystal structure of Cr substituted $ScV_6Sn_6$. Laue diffraction patterns along the $c$-axis for (b) $x = 0.086$ and (c) $x = 0.157$ single crystals. (e) Electronic band structure of $ScV_6Sn_6$ at high temperature (HT) above $T_{CDW}$ and low temperature (LT) in the CDW phase. Green shadows are the possible shifting of Fermi surface. (d) Resistivity as a function of temperature for $x = 0, 0.086, 0.157$ single crystals.

The temperature-dependent longitudinal resistivity ($\rho_{xx}$) for both compositions shows metallic behavior like the parent compound, as displayed in figure 1e. However, the apparent change in $\rho_{xx}$, which indicates the CDW phase transition, systematically varies. The CDW phase transition shifts from 92 K to 50 K for x = 0.086, and vanishes completely for $x = 0.157$, which is consistent with the previous report [26]. The resistivity values at 2 K are typically 3.1 × 10$^{-5}$ Ωcm and 4.5 × 10$^{-5}$ Ωcm, and the residual resistivity ratio ($RRR = \rho_{300K}/\rho_{2K}$) are 3.5 and 3.1 for $x = 0.086$ and 0.157, respectively. It is noteworthy that the substitutions of Y and Lu elements at the Sc position also affect the CDW in a similar manner due to the limitation of rattling in the Sc-Sn-Sn chains in substituted samples [29]. The displacement of Sc and Sn atoms is responsible for strong electron-phonon coupling, the ultimate origin of the CDW [10,20,33,34]. The substitution of Cr atoms for V has two effects: it weakens the bonding and destroys the kagome lattice, and it shifts the Fermi energy ($E_F$) away from the von Hove singularity (vHS) (shown in figure 1d) [26]. Therefore, the CDW phase transition is unlikely to be induced by the FS nesting mechanism, where the vHS is significant, but rather by the strong electron-phonon coupling [20,33,34].

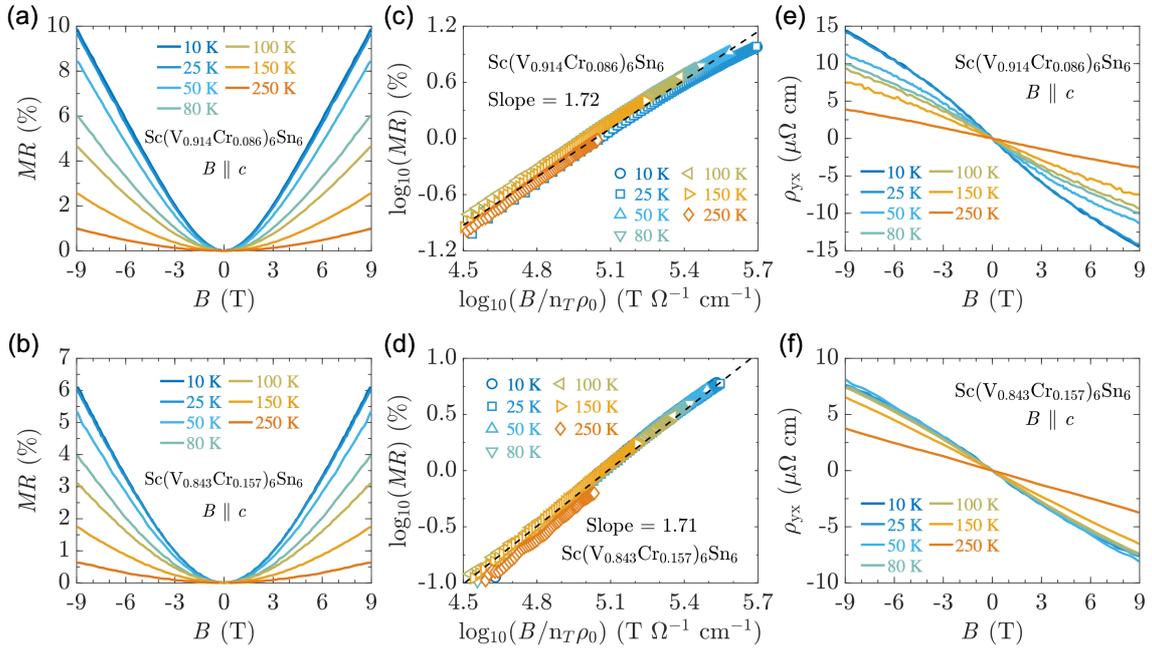

**Figure 2.** Magneto-transport properties of Sc(V$_{1-x}$Cr$_x$)$_6$Sn$_6$. Magnetoresistance at various temperatures for (a) $x = 0.086$ and (b) $x = 0.157$. Extended Kohle's scaling plots for (c) $x = 0.086$ and (d) $x = 0.157$. Their corresponding Hall resistivity in (e) and (f).

To investigate the effect of a magnetic field on $\rho_{xx}$, we measured field-dependent $\rho_{xx}$ of the doped crystals and pressurized pristine ScV$_6$Sn$_6$ crystal at various temperatures. The estimated magnetoresistance is shown in figures 2a and 2b, where *MR* is defines as $[\rho_{xx}(B) - \rho_{xx}(0)/\rho_{xx}(0)] \times 100\%$. $\rho_{xx}(B)$ and $\rho_{xx}(0)$ are the resistivities with and without a field, respectively. The *MR* is significantly reduced to 10% for $x = 0.086$ and 6 % for $x = 0.157$, which is almost 20 and 30 times lower, respectively, than that of the pristine compound [30]. This reduction can be attributed to the shift of the $E_F$ away from the Dirac band and vHS after substituting the Cr [26], resulting in the domination of the trivial charge carriers. Furthermore, the *MR* behavior can be described by the Kohler's law, where $MR \propto (B/\rho_{xx}(0))^m$, and *m* is a temperature independent constant. When single scattering dominates at all temperatures, the Kohler's law allows the *MR* curves of all temperatures to collapse onto a single line. However, the role of temperature dependent charge carrier was recognized later, and then the extended Kohler's law is used [35]. This law describes the $MR \propto (B/n_T \rho_{xx}(0))^m$, where the temperature dependent term $n_T = e\, (\sum_i n_i\mu_i)^{3/2}/(\sum_i n_i\mu_i^3)^{1/2}$ is included as a temperature correlation. The expression of $n_T$ depends on the carrier concentration ($n_i$) and mobility ($\mu_i$) extracted from the Hall resistivity. Figures 2c and 2d show the extended Kohler's plots for *x* = 0.086 and 0.157. The plot for different temperatures at *x* = 0.086 almost coincides in the log-log plot, yielding *m* = 1.72. However, the curves slightly deviate from the line below 50 K from the line due to the CDW transition, which is consistent with the pristine ScV$_6$Sn$_6$ [36]. The deviation in the plot is due to the evolution of electronic band structure at CDW transition. The plot for *x* = 0.157, as displayed in figure 2d, also shows linear dependence in $B/n_T\rho_0$ and yields *m* = 1.71 without any deviation, indicating the complete suppression of CDW. Figures 2e and 2f display the Hall resistivity $\rho_{yx}$ of *x* = 0.086 and 0.157 at several temperatures, respectively. The $\rho_{yx}$ exhibits a nonlinear dependence on the magnetic field at low temperatures. After substituting Cr, the $\rho_{yx}$ changes remarkably compared to the parent compound [30,31]. We estimated the conductivity $\sigma_{xx} = \frac{\rho_{xx}}{\rho_{xx}^2+\rho_{yx}^2}$ and Hall conductivity $\sigma_{xy} = \frac{\rho_{yx}}{\rho_{xx}^2+\rho_{yx}^2}$, as shown in figures 3a and 3b. The two-band model can simultaneously reproduce $\sigma_{xx}$ and $\sigma_{xy}$, yielding two electron-type carriers with different mobility and density [32,37–39]. Figures 3c-3d and 3e-3f show the temperature dependent carrier concentrations and mobilities of both type of electrons for *x* = 0.086 and 0.157. The carrier concentration and mobility display a monotonous relation with temperature without any significant change around $T_{CDW}$ = 50 K for $x$ = 0.086. This result differs from the two-band model in the pristine compound, where a kink appears at CDW transition [31]. One type of electrons, named electron-1, exhibit weak temperature dependence with the density of $10^{19}$ cm$^{-3}$ and a mobility of $10^3$ cm$^2$ V$^{-1}$ s$^{-1}$. However, the density of electron-2 decreases from $10^{22}$ cm$^{-3}$ to $10^{20}$ cm$^{-3}$ as the temperature is decreased from 250 K to 2 K. On increasing Cr concentration up to $x$ = 0.157, the carrier concentration and mobility show similar temperature dependence as $x$ = 0.086 with closely related values. To understand the origin of the Hall resistivity in the Cr substituted sample, we compare it with the pristine ScV$_6$Sn$_6$ sample. Previous reports on pristine crystal [30,31] has shown anomalous Hall effect (AHE)-like behavior in the Hall resistivity, which may be due to the formation of a loop current in the V kagome lattice, breaking the time reversal symmetry (TRS) [17,27]. However, substitution of the Cr atoms in V site in the kagome lattice destroys the V-V bonds by inducing disorder. The presence of an additional valence electron in Cr causes the $E_F$ to shift upwards by 120 meV [26]. This shift also moves the vHS further away from the $E_F$, which is the main cause of the loop current order in AV$_3$Sb$_5$ systems [40–43].

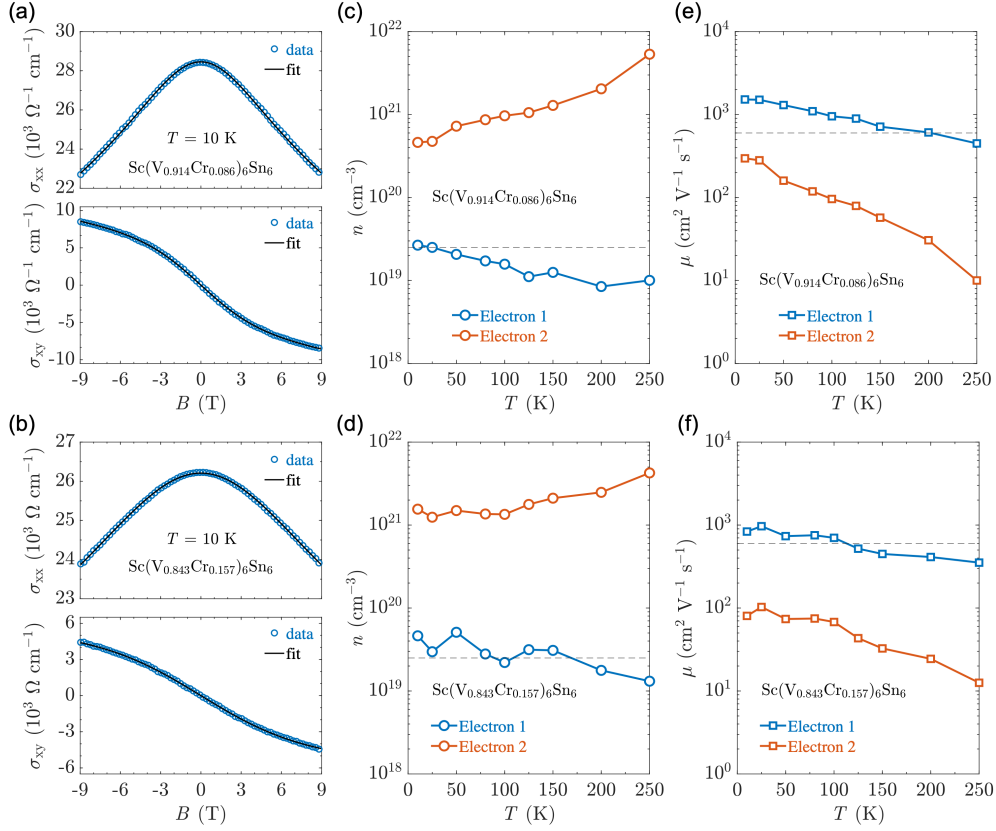

**Figure 3.** Two-band model analysis of Hall resistivity of Sc(V$_{1-x}$Cr$_x$)$_6$Sn$_6$. Magnetic field dependent longitudinal conductivity and Hall conductivity at 10 K for (a) $x$ = 0.086 and $x$ = 0.157. Estimated temperature-dependent carrier concentration (c) and (d), mobility (e) and (f) for $x$ = 0.086, 0.157, respectively.

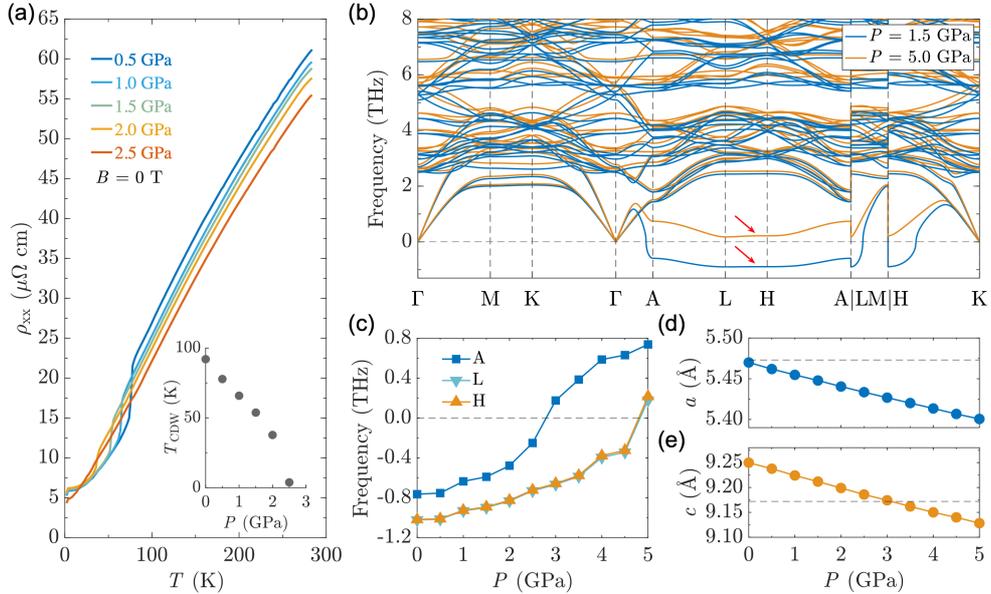

**Figure 4.** Hydrostatic pressure-tuned charge density wave in parent compound ScV$_6$Sn$_6$. (a) Temperature-dependent longitudinal resistivity at several hydrostatic pressures. The inset is the change in CDW transition with pressure. (b) Calculated phonon band dispersions of two selected pressure, 1.5 and 5.0 GPa. (c) Pressure dependent imaginary phonon frequency at high

symmetry *A*, *L*, and *H* points of (b). (d) and (e) Completely relaxed lattice parameters from calculation as the function of pressure. The dashed lines indicate real values of lattice parameters of ScV$_6$Sn$_6$ at ambient pressure.

After studying the chemical substitution effect in the ScV$_6$Sn$_6$ crystals, we investigated the impact of external pressure effect on the electromagnetic transport properties and phonon band structures of pristine ScV$_6$Sn$_6$ crystal. Our external hydrostatic pressure measurements on the parent compound, as shown in figure 4a, demonstrate a systematic suppression of the CDW with pressure, as similar to the chemical substitution. At 1.5 GPa, the CDW appears at 54 K, which eventually disappears at 2.5 GPa, similar to the findings of a previous report [28]. The pressure-temperature phase diagram, shown in the inset of figure 4a, illustrates the pressure-dependent CDW. The rate of change of $T_{CDW}$ with pressure is slower than that of AV$_3$Sb$_5$ systems [44–46]. The vanishing of CDW due to external pressure may suggest the similar physics, as observed with the chemical substitution. Additionally, first-principles calculations using density functional theory (DFT (see method section) were carried out to examine the dynamic instability of ScV$_6$Sn$_6$ under pressure. The implementation of generalized gradient approximation (GGA) method typically overestimates both pressures and relaxed lattice constants in numerical calculations [47,48]. Figure 4b shows the phonon spectrum at pressures of 1.5 and 5 GPa. It is evident that the entire spectrum shifts upward with compressed volume due to hydrostatic pressure. Remarkably, the nearly flat imaginary modes at the $k_z = \pi$ plane at lower pressure, such as 1.5 GPa, are renormalized at higher pressure 5 GPa, indicating the stabilization of the pristine structure under higher pressure. The frequencies at *A*, *L*, and *H* points were tracked under various pressures, as shown in figure 4c, suggesting a critical pressure of ~5 GPa and indicating a stabilized state under this pressure. The completely relaxed lattice constants are also summarized in figures 4d and 4e. Taking into account the temperature effect and overestimation of the lattice constants, the calculated result of approximately 5 GPa is in good agreement with the experimental value of 2.5 GPa.

We have now studied the evolution of *MR* of parent compound with pressure. The *MR* shows a quasilinear behavior with magnetic field without any remarkable change up to 2 GPa as shown in figure 5a. However, at the critical pressure of 2.5 GPa, the quasilinear behavior turns to linear behavior. Conversely, the temperature evolution of *MR* at a fixed pressure does not change (figure 5b) and it is similar to that without pressure [30]. At a pressure of 0.5 GPa, the Shubnikov de Haas oscillations are observed at a high magnetic field range as shown in figure S3 in SM [32]. After analyzing the quantum oscillation, two frequencies are observed, corresponding to 25 T and 47 T, which are consistent with the results obtained at ambient pressure [30,32]. A study of the pressure dependence of these oscillations can provide further insight into changes of the FS. Interestingly, our measured nonlinear Hall resistivity (figure 5c) exhibits a rapid increase with magnetic field and then saturates at around 2 T, which is a striking feature similar to the AHE behavior in soft ferromagnets [49,50]. At the particular pressure of 1.5 GPa, a sign change in Hall signal appears at $T_{CDW} \sim 54$ K, as displayed in figure 5d (see figure S5 for the Hall resistivity of the other pressures in SM) [32]. The Hall resistivity data at 5 K and 80 K, which are above and above $T_{CDW} \sim 54$ K, respectively, were carefully fitted by using a two-band model (see figure S4b in SM) [32]. The data roughly fit and give unreliable fitting parameters, as similar results of ambient pressure [30,32]. Meanwhile, no sign change of the carrier is observed from two-band model as expected from the Hall resistivity, indicating that the nonlinear Hall effect is possibly an AHE.

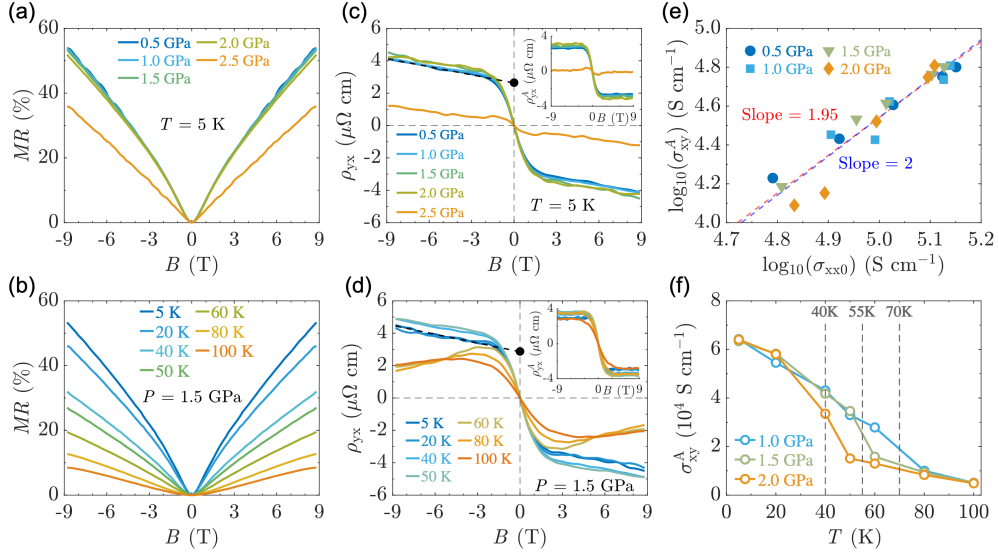

**Figure 5.** Hydrostatic pressure tuned magnetoresistance (MR) and Hall resistivity of ScV$_6$Sn$_6$. (a) MR at $T$ =5 K for several pressures. (b) MR at 1.5 GPa for several temperatures. (c) Hall resistivity at $T$ =5 K for several pressures. Inset shows anomalous Hall resistivity (AHE) after subtracting linear part at high field. (d) Hall resistivity at 1.5 GPa for several temperatures. Inset shows AHE after subtracting linear part at high field. (e) Scaling relationship between anomalous Hall conductivity (AHC) and zero-field longitudinal conductivity at different pressures. (f) AHC as a function of temperature at selective pressures.

The extracted anomalous Hall resistivity at several temperatures and pressures is displayed in the insets in figures 5c and 5d. The longitudinal and anomalous Hall conductivities (AHC) were calculated by using $\sigma_{xx} = \frac{\rho_{xx}}{\rho_{xx}^2 + \rho_{yx}^2}$ and $\sigma_{xy}^A = \frac{\rho_{yx}^A}{\rho_{xx}^2 + \rho_{yx}^{A\,2}}$, respectively, where $\rho_{xx} = \rho_{yy}$ is assumed for simplicity. The values of $\sigma_{xx}$ and $\sigma_{xy}^A$ are typically found to be $10^5$ $\Omega^{-1}$ cm$^{-1}$ and $10^4$ $\Omega^{-1}$ cm$^{-1}$, respectively. To understand the origin of the AHE like behavior from the unified theory, three conductivity regimes are normally identified [51,52]. (i) The high conductivity regime ($\sigma_{xx} > 10^6$ $\Omega^{-1}$ cm$^{-1}$) where the $\sigma_{xy}^A$ follows $\sigma_{xy}^A \propto \sigma_{xx}$, (ii) the intrinsic regime ($10^4$ $\Omega^{-1}$ cm$^{-1}$ < $\sigma_{xx}$ < $10^6$ $\Omega^{-1}$ cm$^{-1}$) where $\sigma_{xy}^A$ is a constant and it is contributed by the Berry curvature, (iii) the low conducting hopping regime ($\sigma_{xx} < 10^6$ $\Omega^{-1}$ cm$^{-1}$) where $\sigma_{xy}^A$ follows a power law relation with $\sigma_{xy}^A \propto \sigma_{xx}^{1.6}$. Our pressure-dependent $\sigma_{xy}^A$ exhibits a power law relation with an exponent close to 2, as displayed in figure 5e. The observed exponent of ~2 goes beyond the above-mentioned empirical regimes. However, the range of $\sigma_{xx}$ (= $10^5$ $\Omega^{-1}$

cm$^{-1}$) lies in the region (ii), indicating an intrinsic origin of the AHE. Meanwhile, figure 5f shows the temperature-dependent evolution of AHE-like behavior at different pressures. At each pressure, the AHC monotonically decreases up to its respective $T_{CDW}$ and declines rapidly above it. For example, the slope changes at 50 K in 1.5 GPa and at 35 K in 2 GPa, which are quite similar to the observation at ambient pressure at 92 K [30]. This unambiguously indicates that the AHE-like behavior is closely related to the CDW phase and needs further exploration.

The absence of any magnetic order in ScV$_6$Sn$_6$ makes the AHE-like behavior intriguing and opens up further discussion. Both the chemical substitution and pressure tune the Sc–Sn and Sn–Sn bonds distance [26,28,29], leading to the renormalization of imaginary phonon modes and suppression of the CDW. However, the pressures below 2 GPa do not significantly affect the behavior of Hall resistivity, indicating a marginal effect on the electronic instability. Once CDW disappears, the AHE-like Hall signal also disappears. The AV$_3$Sb$_5$ family has also shown AHE-like behavior, with various speculations made, including the concept of loop current order [40–42,53–56]. A very recent study of muon-spin spectroscopy (μSR) measurements on RbV$_3$Sb$_5$ shows a clear distinction at the CDW transition. The μSR signal enhances in thin layer samples and diminishes after doping in the kagome lattice [57]. In ScV$_6$Sn$_6$, μSR spectroscopy has uncovered the hidden magnetism, with the μSR signal increasing with field [27]. The loss of AHE-like feature in Cr-substituted samples strengthens the claim of hidden magnetism in the form of the loop current.

**Conclusion**

In summary, we have presented a comprehensive study of temperature-dependent magnetoelectrical properties of the kagome metal ScV$_6$Sn$_6$ through chemical substitution and hydrostatic pressure. Our results show that the CDW phase transition decreases with either substituting Cr at the V site or applying pressure. For substituted samples, the quadratic MR follows the extended Kohle's scaling law, and the Hall resistivity can be well described by the two-band model. In contrast, the parent compound under pressure retains the quasilinear MR and AHE-like behavior, which declines rapidly above the corresponding $T_{CDW}$ for each pressure. The observed AHE-like behavior seems to result from hidden magnetism that breaks TRS. Our findings light a path for the study of unconventional electronic properties in kagome metallic materials under external stimuli.

**Methods**

**Crystal growth**

Single crystals of Sc(V$_{1-x}$Cr$_x$)$_6$Sn$_6$ (x = 0, 0.05, 0.157) are grown by Sn-flux method following the previous report [ref PRL, PRB]. High-purity Sc, V, Cr, and Sn elements are cut into small pieces, loaded into an alumina crucible, and then sealed into a high-evacuated silica tube. The tube then was heated to 1327 K, maintained for 20 hours and then slowly cooled down to 1073 K at a ratio of 2 K/h. The hexagon-shaped single crystals are yielded on the bottom of crucible after centrifuge.

**Crystal characterization**

The x-ray diffraction (XRD) data is collected in a home-made x-ray diffractometer equipped with $Cu\ K_{\alpha 1}$ radiation ($\lambda \sim 1.54$ Å). Crystal structure is refined via FullProf package [58]. The chemical components are confirmed by EDXS option on a commercial scanning electronic microscopy. The Laue backscattering diffraction pattern are collected in a home-made Laue diffractometer to confirm the crystal orientation.

**Transport properties at ambient and hydrostatic pressure**

The crystals are cut into cuboid along specific crystallographic orientation to facilitate the measurement of magnetoelectrical transport properties. The longitudinal and Hall measurements for doped crystals were utilized in Physical Properties Measurement System (PPMS, Quantum Design Inc.). For the measurement under pressure, a standard piston-cylinder hydrostatic pressure cell (Quantum Design Inc.) was used in PPMS. The pressure was simultaneously calibrated by measuring the superconducting critical temperature shift of Pb element. A six probes method for all measurements in ambient pressure for doped crystals and high pressure for pristine crystals.

**Phonon band calculation**

The first-principles calculations are performed using the density functional theory (DFT) and density functional perturbation theory (DFPT), based on the generalized gradient approximation (GGA) with Perbew-Burke-Ernzerhof (PBE) exchange-correlation functional as implemented in Vienna ab-initio Simulation Package (VASP) with the projector augmented wave (PAW) pseudopotentials [59–62]. The structures were fully relaxed with a force convergence criterion of $10^{-3}$ eV/Å under pressure. The phonon spectra are calculated based on $2 \times 2 \times 2$ supercells with a Γ-centered $4 \times 4 \times 2$ $k$-mesh with an energy cutoff of 450 eV. PHONOPY code was utilized to extract the force constants and phonon spectra [63]. The spin-orbital coupling (SOC) was not included in the phonon calculations. The discrepancy between the relaxed lattice constants and experimental values could be caused by the pseudopotentials and GGA.

**Acknowledgements**

This work was financially supported by the Deutsche Forschungsgemeinschaft (DFG) under SFB1143 (project no. 247310070), the Würzburg-Dresden Cluster of Excellence on Complexity and Topology in Quantum Matter—ct.qmat (EXC 2147, project no. 390858490) and the QUAST-FOR5249-449872909. NK acknowledges Max Planck Society for funding support under Max Planck-India partner group project and SERB for financial support through Grant Sanction No. CRG/2021/002747.

**References**

[1] W.-S. Wang, Z.-Z. Li, Y.-Y. Xiang, and Q.-H. Wang, *Competing Electronic Orders on Kagome Lattices at van Hove Filling*, Phys. Rev. B **87**, 11 (2013).

[2] M. Li, Q. Wang, G. Wang, Z. Yuan, W. Song, R. Lou, Z. Liu, Y. Huang, Z. Liu, H. Lei, et al., *Dirac Cone, Flat Band and Saddle Point in Kagome Magnet YMn$_6$Sn$_6$*, Nat Commun **12**, 1 (2021).

[3] Y. Hu, X. Wu, Y. Yang, S. Gao, N. C. Plumb, A. P. Schnyder, W. Xie, J. Ma, and M. Shi, *Tunable Topological Dirac Surface States and van Hove Singularities in Kagome Metal GdV$_6$Sn$_6$*, Sci. Adv. **8**, 38 (2022).

[4] D. Zhang, Z. Hou, and W. Mi, *Progress in Magnetic Alloys with Kagome Structure: Materials, Fabrications and Physical Properties*, J. Mater. Chem. C **10**, 20 (2022).

[5] Y. Hu, X. Wu, A. P. Schnyder, and M. Shi, *Electronic Landscape of Kagome Superconductors AV$_3$Sb$_5$ (A = K, Rb, Cs) from Angle-Resolved Photoemission Spectroscopy*, Npj Quantum Mater. **8**, 1 (2023).

[6] Y. Wang, H. Wu, G. T. McCandless, J. Y. Chan, and M. N. Ali, *Quantum States and Intertwining Phases in Kagome Materials*, Nat Rev Phys **5**, 11 (2023).

[7] X.-W. Yi, Z.-W. Liao, J.-Y. You, B. Gu, and G. Su, *Superconducting, Topological, and Transport Properties of Kagome Metals CsTi$_3$Bi$_5$ and RbTi$_3$Bi$_5$*, Research **6**, 0238 (2023).

[8] B. R. Ortiz, L. C. Gomes, J. R. Morey, M. Winiarski, M. Bordelon, J. S. Mangum, I. W. H. Oswald, J. A. Rodriguez-Rivera, J. R. Neilson, S. D. Wilson, et al., *New Kagome Prototype Materials: Discovery of KV$_3$Sb$_5$, RbV$_3$Sb$_5$, and CsV$_3$Sb$_5$*, Phys. Rev. Materials **3**, 9 (2019).

[9] X. Teng, L. Chen, F. Ye, E. Rosenberg, Z. Liu, J.-X. Yin, Y.-X. Jiang, J. S. Oh, M. Z. Hasan, K. J. Neubauer, et al., *Discovery of Charge Density Wave in a Kagome Lattice Antiferromagnet*, Nature **609**, 490 (2022).

[10] H. W. S. Arachchige, W. R. Meier, M. Marshall, T. Matsuoka, R. Xue, M. A. McGuire, R. P. Hermann, H. Cao, and D. Mandrus, *Charge Density Wave in Kagome Lattice Intermetallic ScV$_6$Sn$_6$*, Phys. Rev. Lett. **129**, 21 (2022).

[11] H. Tan, Y. Liu, Z. Wang, and B. Yan, *Charge Density Waves and Electronic Properties of Superconducting Kagome Metals*, Phys. Rev. Lett. **127**, 4 (2021).

[12] L. Nie, K. Sun, W. Ma, D. Song, L. Zheng, Z. Liang, P. Wu, F. Yu, J. Li, M. Shan, et al., *Charge-Density-Wave-Driven Electronic Nematicity in a Kagome Superconductor*, Nature **604**, 7904 (2022).

[13] Q. Stahl, D. Chen, T. Ritschel, C. Shekhar, E. Sadrollahi, M. C. Rahn, O. Ivashko, M. V. Zimmermann, C. Felser, and J. Geck, *Temperature-Driven Reorganization of Electronic Order in CsV$_3$Sb$_5$*, Phys. Rev. B **105**, 19 (2022).

[14] R. Tazai, Y. Yamakawa, S. Onari, and H. Kontani, *Mechanism of Exotic Density-Wave and beyond-Migdal Unconventional Superconductivity in Kagome Metal AV$_3$Sb$_5$ (A = K, Rb, Cs)*, Sci. Adv. **8**, 13 (2022).

[15] L. Zheng, Z. Wu, Y. Yang, L. Nie, M. Shan, K. Sun, D. Song, F. Yu, J. Li, D. Zhao, et al., *Emergent Charge Order in Pressurized Kagome Superconductor CsV$_3$Sb$_5$*, Nature **611**, 7937 (2022).

[16] D. R. Saykin, C. Farhang, E. D. Kountz, D. Chen, B. R. Ortiz, C. Shekhar, C. Felser, S. D. Wilson, R. Thomale, J. Xia, et al., *High Resolution Polar Kerr Effect Studies of CsV$_3$Sb$_5$: Tests for Time-Reversal Symmetry Breaking below the Charge-Order Transition*, Phys. Rev. Lett. **131**, 1 (2023).

[17] T. Asaba, A. Onishi, Y. Kageyama, T. Kiyosue, K. Ohtsuka, S. Suetsugu, Y. Kohsaka, T. Gaggl, Y. Kasahara, H. Murayama, et al., *Evidence for an Odd-Parity Nematic Phase above the Charge-Density-Wave Transition in a Kagome Metal*, Nat. Phys. **20**, 1 (2024).

[18] X. Teng, J. S. Oh, H. Tan, L. Chen, J. Huang, B. Gao, J.-X. Yin, J.-H. Chu, M. Hashimoto, D. Lu, et al., *Magnetism and Charge Density Wave Order in Kagome FeGe*, Nat. Phys. **19**, 6 (2023).

[19] Y. Wang, *Enhanced Spin-Polarization via Partial Ge-Dimerization as the Driving Force of the Charge Density Wave in FeGe*, Phys. Rev. Materials **7**, 10 (2023).

[20] A. Korshunov, H. Hu, D. Subires, Y. Jiang, D. Călugăru, X. Feng, A. Rajapitamahuni, C. Yi, S. Roychowdhury, M. G. Vergniory, et al., *Softening of a Flat Phonon Mode in the Kagome $ScV_6Sn_6$*, Nat Commun **14**, 1 (2023).

[21] Y. Hu, J. Ma, Y. Li, D. J. Gawryluk, T. Hu, J. Teyssier, V. Multian, Z. Yin, Y. Jiang, S. Xu, et al., *Phonon Promoted Charge Density Wave in Topological Kagome Metal $ScV_6Sn_6$*, arXiv:2304.06431.

[22] Y. Gu, E. T. Ritz, W. R. Meier, A. Blockmon, K. Smith, R. P. Madhogaria, S. Mozaffari, D. Mandrus, T. Birol, and J. L. Musfeldt, *Phonon Mixing in the Charge Density Wave State of $ScV_6Sn_6$*, Npj Quantum Mater. **8**, 1 (2023).

[23] T. Hu, H. Pi, S. Xu, L. Yue, Q. Wu, Q. Liu, S. Zhang, R. Li, X. Zhou, J. Yuan, et al., *Optical Spectroscopy and Band Structure Calculations of the Structural Phase Transition in the Vanadium-Based Kagome Metal $ScV_6Sn_6$*, Phys. Rev. B **107**, 16 (2023).

[24] D. Di Sante, C. Bigi, P. Eck, S. Enzner, A. Consiglio, G. Pokharel, P. Carrara, P. Orgiani, V. Polewczyk, J. Fujii, et al., *Flat Band Separation and Robust Spin Berry Curvature in Bilayer Kagome Metals*, Nat. Phys. **19**, 8 (2023).

[25] S. Cheng, Z. Ren, H. Li, J. Oh, H. Tan, G. Pokharel, J. M. DeStefano, E. Rosenberg, Y. Guo, Y. Zhang, et al., *Nanoscale Visualization and Spectral Fingerprints of the Charge Order in ScV6Sn6 Distinct from Other Kagome Metals*, (2023).

[26] S. Lee, C. Won, J. Kim, J. Yoo, S. Park, J. Denlinger, C. Jozwiak, A. Bostwick, E. Rotenberg, R. Comin, et al., *Nature of Charge Density Wave in Kagome Metal $ScV_6Sn_6$*, Npj Quantum Mater. **9**, 15 (2024).

[27] Z. Guguchia, D. J. Gawryluk, S. Shin, Z. Hao, C. Mielke Iii, D. Das, I. Plokhikh, L. Liborio, J. K. Shenton, Y. Hu, et al., *Hidden Magnetism Uncovered in a Charge Ordered Bilayer Kagome Material $ScV_6Sn_6$*, Nat Commun **14**, 1 (2023).

[28] X. Zhang, J. Hou, W. Xia, Z. Xu, P. Yang, A. Wang, Z. Liu, J. Shen, H. Zhang, X. Dong, et al., *Destabilization of the Charge Density Wave and the Absence of Superconductivity in $ScV_6Sn_6$ under High Pressures up to 11 GPa*, Materials **15**, 20 (2022).

[29] W. R. Meier, R. P. Madhogaria, S. Mozaffari, M. Marshall, D. E. Graf, M. A. McGuire, H. W. S. Arachchige, C. L. Allen, J. Driver, H. Cao, et al., *Tiny Sc Allows the Chains to Rattle: Impact of Lu and Y Doping on the Charge-Density Wave in $ScV_6Sn_6$*, J. Am. Chem. Soc. **145**, 38 (2023).

[30] C. Yi, X. Feng, N. Mao, P. Yanda, S. Roychowdhury, Y. Zhang, C. Felser, and C. Shekhar, *Quantum Oscillations Revealing Topological Band in Kagome Metal $ScV_6Sn_6$*, Phys. Rev. B **109**, 3 (2024).

[31] S. Mozaffari, W. R. Meier, R. P. Madhogaria, N. Peshcherenko, S.-H. Kang, J. W. Villanova, H. W. S. Arachchige, G. Zheng, Y. Zhu, K.-W. Chen, et al., *Universal Sublinear Resistivity in Vanadium Kagome Materials Hosting Charge Density Waves*, arXiv:2305.02393.

[32] Supplementary material at XX XX XXX XXX includes: crystalline structure and chemical components characterization, quantum oscillation analysis, and additional two-band model fitting of Cr substituted samples and pressurized samples., (n.d.).

[33] S. Cao, C. Xu, H. Fukui, T. Manjo, Y. Dong, M. Shi, Y. Liu, C. Cao, and Y. Song, *Competing Charge-Density Wave Instabilities in the Kagome Metal $ScV_6Sn_6$*, Nat Commun **14**, 1 (2023).

[34] H. Tan and B. Yan, *Abundant Lattice Instability in Kagome Metal ScV$_6$Sn$_6$*, Phys. Rev. Lett. **130**, 26 (2023).

[35] J. Xu, F. Han, T.-T. Wang, L. R. Thoutam, S. E. Pate, M. Li, X. Zhang, Y.-L. Wang, R. Fotovat, U. Welp, et al., *Extended Kohler's Rule of Magnetoresistance*, Phys. Rev. X **11**, 4 (2021).

[36] J. M. DeStefano, E. Rosenberg, O. Peek, Y. Lee, Z. Liu, Q. Jiang, L. Ke, and J.-H. Chu, *Pseudogap Behavior in Charge Density Wave Kagome Material ScV$_6$Sn$_6$ Revealed by Magnetotransport Measurements*, Npj Quantum Mater. **8**, 65 (2023).

[37] Y. L. Wang, L. R. Thoutam, Z. L. Xiao, J. Hu, S. Das, Z. Q. Mao, J. Wei, R. Divan, A. Luican-Mayer, G. W. Crabtree, et al., *Origin of the Turn-on Temperature Behavior in WTe$_2$*, Phys. Rev. B **92**, 180402 (2015).

[38] F. C. Chen, Y. Fei, S. J. Li, Q. Wang, X. Luo, J. Yan, W. J. Lu, P. Tong, W. H. Song, X. B. Zhu, et al., *Temperature-Induced Lifshitz Transition and Possible Excitonic Instability in ZrSiSe*, Phys. Rev. Lett. **124**, 236601 (2020).

[39] J. Xu, Y. Wang, S. E. Pate, Y. Zhu, Z. Mao, X. Zhang, X. Zhou, U. Welp, W.-K. Kwok, D. Y. Chung, et al., *Unreliability of Two-Band Model Analysis of Magnetoresistivities in Unveiling Temperature-Driven Lifshitz Transition*, Phys. Rev. B **107**, 035104 (2023).

[40] H. Li, Y. B. Kim, and H.-Y. Kee, *Intertwined Van-Hove Singularities as a Mechanism for Loop Current Order in Kagome Metals*, arXiv:2309.03288.

[41] J.-W. Dong, Z. Wang, and S. Zhou, *Loop-Current Charge Density Wave Driven by Long-Range Coulomb Repulsion on the Kagomé Lattice*, Phys. Rev. B **107**, 4 (2023).

[42] M. H. Christensen, T. Birol, B. M. Andersen, and R. M. Fernandes, *Loop Currents in AV$_3$Sb$_5$ Kagome Metals: Multipolar and Toroidal Magnetic Orders*, Phys. Rev. B **106**, 14 (2022).

[43] R. Tazai, Y. Yamakawa, and H. Kontani, *Charge-Loop Current Order and Z$_3$ Nematicity Mediated by Bond Order Fluctuations in Kagome Metals*, Nat Commun **14**, 1 (2023).

[44] F. Du, S. Luo, B. R. Ortiz, Y. Chen, W. Duan, D. Zhang, X. Lu, S. D. Wilson, Y. Song, and H. Yuan, *Pressure-Induced Double Superconducting Domes and Charge Instability in the Kagome Metal CsV$_3$Sb$_5$*, Phys. Rev. B **103**, L220504 (2021).

[45] K. Y. Chen, N. N. Wang, Q. W. Yin, Y. H. Gu, K. Jiang, Z. J. Tu, C. S. Gong, Y. Uwatoko, J. P. Sun, H. C. Lei, et al., *Double Superconducting Dome and Triple Enhancement of T$_C$ in the Kagome Superconductor CsV$_3$Sb$_5$ under High Pressure*, Phys. Rev. Lett. **126**, 247001 (2021).

[46] N. N. Wang, K. Y. Chen, Q. W. Yin, Y. N. N. Ma, B. Y. Pan, X. Yang, X. Y. Ji, S. L. Wu, P. F. Shan, S. X. Xu, et al., *Competition between Charge-Density-Wave and Superconductivity in the Kagome Metal RbV$_3$Sb$_5$*, Phys. Rev. Research **3**, 043018 (2021).

[47] Y. Li, L. J. Zhang, T. Cui, Y. W. Li, Y. Wang, Y. M. Ma, and G. T. Zou, *First-Principles Studies of Phonon Instabilities in AgI under High Pressure*, J. Phys.: Condens. Matter **20**, 195218 (2008).

[48] M. Catti and L. Di Piazza, *Phase Equilibria and Transition Mechanisms in High-Pressure AgCl by Ab Initio Methods*, J. Phys. Chem. B **110**, 1576 (2006).

[49] K. Manna, L. Muechler, T.-H. Kao, R. Stinshoff, Y. Zhang, J. Gooth, N. Kumar, G. Kreiner, K. Koepernik, R. Car, et al., *From Colossal to Zero: Controlling the Anomalous Hall Effect in Magnetic Heusler Compounds via Berry Curvature Design*, Phys. Rev. X **8**, 041045 (2018).

[50] D. S. Bouma, Z. Chen, B. Zhang, F. Bruni, M. E. Flatté, A. Ceballos, R. Streubel, L.-W. Wang, R. Q. Wu, and F. Hellman, *Itinerant Ferromagnetism and Intrinsic Anomalous Hall Effect in Amorphous Iron-Germanium*, Phys. Rev. B **101**, 014402 (2020).

[51] N. Nagaosa, J. Sinova, S. Onoda, A. H. MacDonald, and N. P. Ong, *Anomalous Hall Effect*, Rev. Mod. Phys. **82**, 1539 (2010).

[52] S.-Y. Yang, Y. Wang, B. R. Ortiz, D. Liu, J. Gayles, E. Derunova, R. Gonzalez-Hernandez, L. Šmejkal, Y. Chen, S. S. P. Parkin, et al., *Giant, Unconventional Anomalous Hall Effect in the Metallic Frustrated Magnet Candidate, $KV_3Sb_5$*, Sci. Adv. **6**, eabb6003 (2020).

[53] C. Mielke, D. Das, J.-X. Yin, H. Liu, R. Gupta, Y.-X. Jiang, M. Medarde, X. Wu, H. C. Lei, J. Chang, et al., *Time-Reversal Symmetry-Breaking Charge Order in a Kagome Superconductor*, Nature **602**, 7896 (2022).

[54] C. Guo, C. Putzke, S. Konyzheva, X. Huang, M. Gutierrez-Amigo, I. Errea, D. Chen, M. G. Vergniory, C. Felser, M. H. Fischer, et al., *Switchable Chiral Transport in Charge-Ordered Kagome Metal $CsV_3Sb_5$*, Nature **611**, 7936 (2022).

[55] X. Feng, K. Jiang, Z. Wang, and J. Hu, *Chiral Flux Phase in the Kagome Superconductor $AV_3Sb_5$*, Science Bulletin **66**, 14 (2021).

[56] L. Yu, C. Wang, Y. Zhang, M. Sander, S. Ni, Z. Lu, S. Ma, Z. Wang, Z. Zhao, H. Chen, et al., *Evidence of a Hidden Flux Phase in the Topological Kagome Metal $CsV_3Sb_5$*, (2021).

[57] J. N. Graham, C. Mielke, D. Das, T. Morresi, V. Sazgari, A. Suter, T. Prokscha, H. Deng, R. Khasanov, S. D. Wilson, et al., *Depth-Dependent Study of Time-Reversal Symmetry-Breaking in the Kagome Superconductor $AV_3Sb_5$*, (2024).

[58] J. Rodríguez-Carvajal, *Recent Advances in Magnetic Structure Determination by Neutron Powder Diffraction*, Physica B: Condensed Matter **192**, 55 (1993).

[59] G. Kresse and J. Furthmüller, *Efficient Iterative Schemes for* Ab Initio *Total-Energy Calculations Using a Plane-Wave Basis Set*, Phys. Rev. B **54**, 11169 (1996).

[60] G. Kresse and J. Hafner, Ab Initio *Molecular-Dynamics Simulation of the Liquid-Metal–Amorphous-Semiconductor Transition in Germanium*, Phys. Rev. B **49**, 14251 (1994).

[61] G. Kresse and J. Furthmüller, *Efficiency of Ab-Initio Total Energy Calculations for Metals and Semiconductors Using a Plane-Wave Basis Set*, Computational Materials Science **6**, 15 (1996).

[62] J. P. Perdew, K. Burke, and M. Ernzerhof, *Generalized Gradient Approximation Made Simple*, Phys. Rev. Lett. **77**, 3865 (1996).

[63] A. Togo and I. Tanaka, *First Principles Phonon Calculations in Materials Science*, Scripta Materialia **108**, 1 (2015).

This supplementary materials for
# Tuning charge density wave of kagome metal ScV$_6$Sn$_6$

Table S1. Contacts geometry for measurements of transport properties.

| Samples | Contacts geometry[*] |
|---|---|
| Sc(V$_{0.914}$Cr$_{0.086}$)$_6$Sn$_6$ | $B \parallel c, I \parallel a$ |
| Sc(V$_{0.843}$Cr$_{0.157}$)$_6$Sn$_6$ | $B \parallel c, I \parallel b$ |
| ScV$_6$Sn$_6$ (pressure) | $B \parallel a, I \parallel b$ |

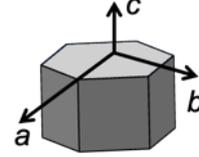

[*] For hexagonal structure, $a$-axis is $[2\,\bar{1}\,\bar{1}\,0]$, $b$-axis is $[0\,1\,\bar{1}\,0]$, and $c$-axis is $[0\,0\,0\,1]$, respectively. During measurement, the applied current direction is defined as $x$ and magnetic field is defined as $z$.

Table S2. Refined lattice parameters of Sc(V$_{1-x}$Cr$_x$)$_6$Sn$_6$ (nominal $x$ = 0, 0.05, 0.1)

| Lattice | ScV$_6$Sn$_6$ | Sc(V$_{0.95}$Cr$_{0.05}$)$_6$Sn$_6$ | Sc(V$_{0.9}$Cr$_{0.1}$)$_6$Sn$_6$ |
|---|---|---|---|
| $a$ (Å) | 5.47303(4) | 5.47019(5) | 5.46950(6) |
| $c$ (Å) | 9.17188(7) | 9.16319(10) | 9.15884(11) |
| $a/c$ | 0.59671845 | 0.596974416 | 0.597182613 |

Table S3. Chemical components determined by EDXS for Sc(V$_{1-x}$Cr$_x$)$_6$Sn$_6$ ($x$ = 0.05, 0.1) samples which are measured in magnetoelectric transport properties.

| | Sc(V$_{0.95}$Cr$_{0.05}$)$_6$Sn$_6$ | | | |
|---|---|---|---|---|
| Elements | Sc | V | Cr | Sn |
| Point1 | 8.15 | 41.45 | 4.12 | 46.28 |
| Point2 | 8.32 | 40.91 | 4.07 | 46.71 |
| Point3 | 8.6 | 40.94 | 3.88 | 46.58 |
| Point4 | 7.67 | 40.76 | 4.2 | 47.37 |
| Point5 | 8.03 | 40.62 | 3.8 | 47.54 |
| Point6 | 8.44 | 41.18 | 3.54 | 46.84 |
| Point7 | 8.31 | 41.21 | 3.73 | 46.75 |
| Point8 | 8.18 | 40.93 | 3.87 | 47.02 |
| Point9 | 8.02 | 40.85 | 3.55 | 47.59 |
| Point10 | 7.95 | 40.91 | 3.63 | 47.52 |
| Average | 8.167 | 40.976 | 3.839 | 47.02 |
| Sc(V$_{0.9}$Cr$_{0.1}$)$_6$Sn$_6$ | | | | |
| Elements | Sc | V | Cr | Sn |
| Point1 | 8.27 | 37.07 | 7.14 | 47.53 |
| Point2 | 8.01 | 37.66 | 6.46 | 47.88 |
| Point3 | 8.41 | 37.86 | 7 | 46.73 |
| Point4 | 8.2 | 37.25 | 7.39 | 47.15 |
| Point5 | 8.53 | 37.98 | 6.73 | 46.77 |
| Point6 | 7.8 | 38.81 | 7.32 | 46.07 |
| Point7 | 8.79 | 37.73 | 6.84 | 46.64 |
| Point8 | 8.18 | 38.21 | 7.14 | 46.47 |
| Point9 | 8.11 | 38.23 | 6.95 | 46.71 |
| Point10 | 8.32 | 37.87 | 7.7 | 46.12 |
| Average | 8.262 | 37.867 | 7.067 | 46.807 |

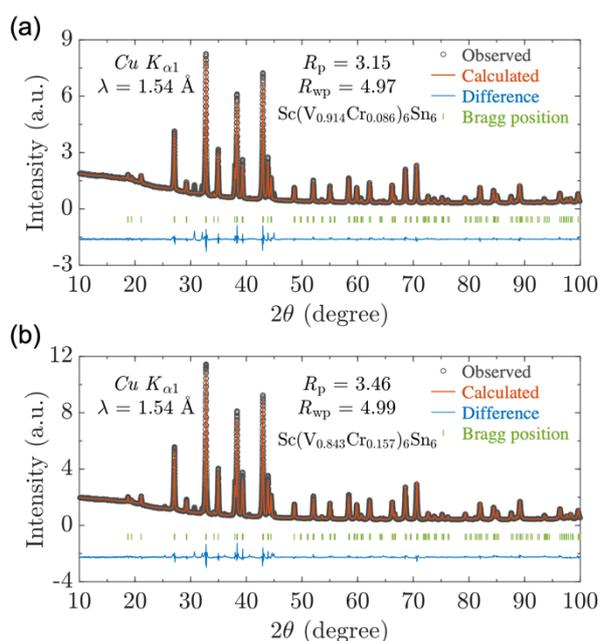

Figure S1. Powder x-ray diffraction for Cr substituted samples for (a) x = 0.086 and (b) x = 0.157, respectively. The refined lattice parameters are summarized in table S2.

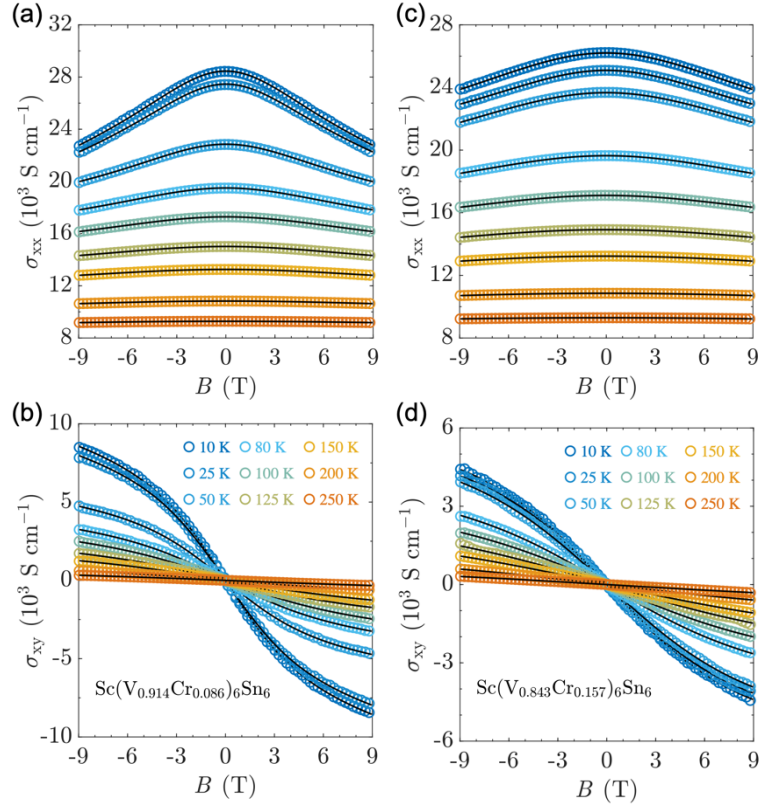

Figure S2. Two-band model analysis of Cr-substituted samples. Simultaneously fitting of (a) longitudinal conductivity and (b) Hall conductivity for x = 0.086. Simultaneously fitting of (c) longitudinal conductivity and (d) Hall conductivity for x = 0.157. The following equations $\sigma_{xx}(B) = e(\frac{n_1\mu_1}{1+\mu_1^2 B^2} + \frac{n_2\mu_2}{1+\mu_2^2 B^2})$ and $\sigma_{xy}(B) = eB(\frac{n_1\mu_1^2}{1+\mu_1^2 B^2} + \frac{n_2\mu_2^2}{1+\mu_2^2 B^2})$ are used for fitting.

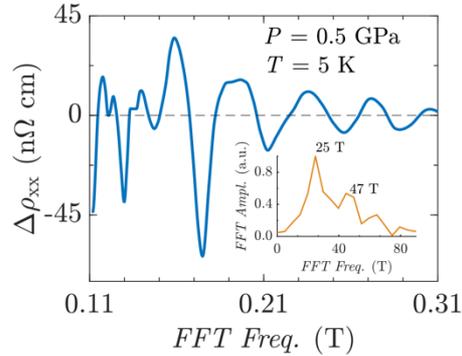

Figure S3. Shubnikov de Haas oscillation observed at P = 0.5 GPa and T = 5 K under high magnetic field range. Inset is the Fast Fourier transform of the oscillation, giving two frequencies.

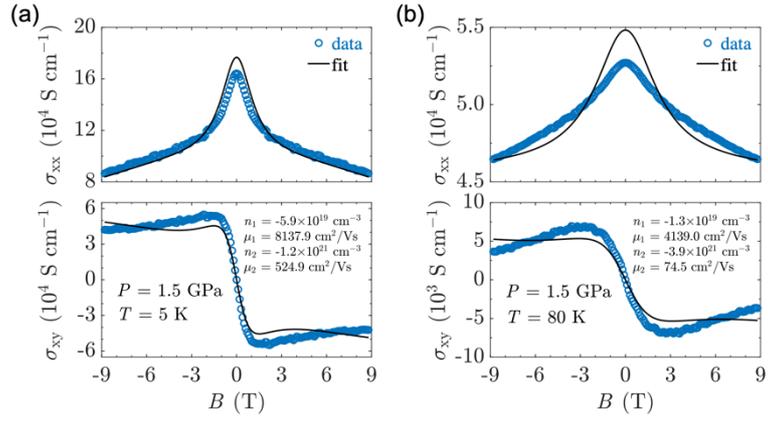

Figure S4. Two-band model analysis of pressurized pristine samples. Simultaneously fitting of (a) longitudinal conductivity and (b) Hall conductivity for $P$ = 1.5 GPa at $T$ = 5 K. Simultaneously fitting of (c) longitudinal conductivity and (d) Hall conductivity for for $P$ = 1.5 GPa at $T$ = 80 K.

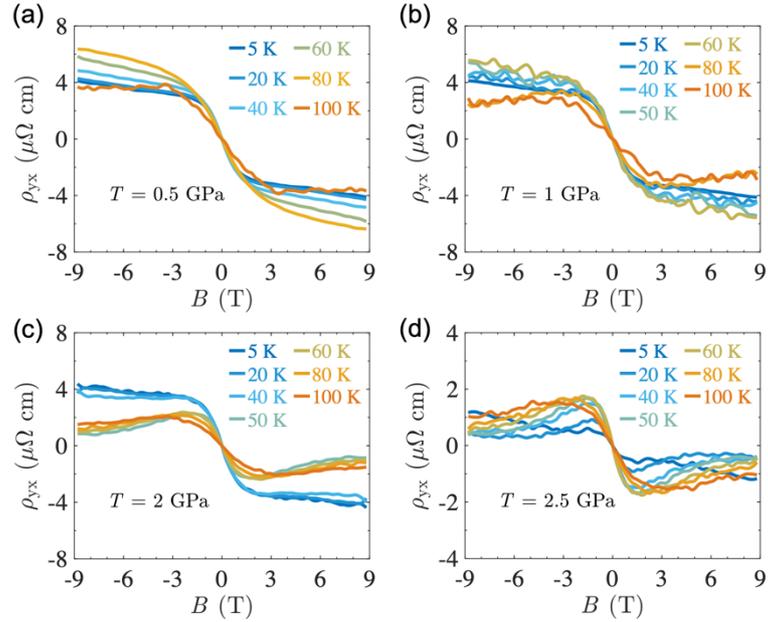

Figure S5. Hall resistivity as function of magnetic field at various temperatures and pressures.